# Energy degrader optimization for medical beam lines

Vladimir Anferov

Indiana University Cyclotron Facility, Bloomington, IN 47408

This paper describes the optimization of a variable energy degrader design for the Midwest Proton Radiotherapy Institute (MPRI) [1]. To optimize the energy degrader design we investigate the choice of an optimal material for the degrader, the beam emittance growth in the degrader, and the matching of the degraded beam with the acceptance of a medical beam line.

## 1. Introduction

Many practical applications of particle accelerators require varying the beam energy in an experimental beam line without changing the settings of the accelerator. Installing a variable thickness energy degrader provides a fast, reliable and reproducible way of setting the beam energy, which is especially important for medical applications. Several proton radiation treatment centers including MPRI [1] and NPTC [2] plan on using variable energy degraders to set the beam energy for radiation treatments. A typical setup includes a variable thickness degrader followed by a momentum selection beam line that transfers the beam to a treatment room. Multiple scattering in the energy degrader causes beam emittance growth and a rotation in phase space, which results in a mismatch of the degraded beam with the acceptance of the momentum selection beam line. In order to reduce beam losses in the energy degrading process it is important to minimize both effects. In this paper we investigate an optimal configuration for the degrader including the choice of material and geometry. An excellent review of various energy degrader configurations in medical beam lines is given in [3]. We consider three generic options for the variable thickness degrader geometry (shown in Figure 1): a wedge with a flat front face, a wedge with a flat exit face, and a wedge with a fixed center of gravity (or double wedge system). This study expands the analysis of the degrader to beam line matching performed at NPTC [4].

## 2. Emittance growth in energy degrader

The beam emittance is determined by the rms size and angular spread of the beam. Both the beam spot size and the angular spread grow due to multiple Coulomb scattering in the degrader material. The lateral displacement of a scattered particle accumulates as an integral of the multiple scattering angle over the degrader thickness. Empirically, it is proportional to the rms multiple scattering angle and to the material thickness $L$ [5]:

$$\sqrt{\langle x_{MS}^2 \rangle} = \frac{L}{\sqrt{3}} \sqrt{\langle \theta_x^2 \rangle} \tag{1}$$

With a typical thickness of an energy degrader being of the order of a few centimeters, the angular scattering is always an order of magnitude larger than the lateral scattering, and therefore, its contribution dominates in the emittance growth. The effect of the multiple scattering to the beam emittance is illustrated in Figure 2. The rms multiple scattering angle in one plane is given by [5]:

$$\sqrt{\langle \theta_x^2 \rangle} = \frac{z \cdot 13.6 \cdot MeV}{\beta c \cdot p} \sqrt{\frac{L}{L0}} \cdot \left(1 + 0.038 \cdot \ln\left(\frac{L}{L0}\right)\right) \tag{2}$$

where $\beta c$, $z$ and $p$ are velocity, charge and momentum of the incident particle and $L0$ is the radiation length in the degrader material of thickness $L$. Multiple scattering is statistically independent from the beam initial conditions. Therefore, multiple scattering angle should be added in quadrature to the rms beam angular spread

$$\langle x_1'^2 \rangle = \langle x_0'^2 \rangle + \langle \theta_x^2 \rangle \tag{3}$$

To express the beam emittance growth due to multiple scattering we will use the following emittance parameterization:

$$\varepsilon = \gamma \cdot x^2 + 2\alpha \cdot x' x + \beta \cdot x'^2 \tag{4}$$

where $\alpha$, $\beta$, and $\gamma$ are the optical functions describing beam envelope evolution along the beam line. Using equations (3) and (4) and neglecting the contribution of the lateral scattering, one can conclude that the beam emittance increase after a thin degrader is proportional to the beta-function and the mean square multiple scattering angle:

$$\varepsilon_1 = \varepsilon_0 + \beta \langle \theta_x^2 \rangle \qquad (5)$$

The beta-function describes the beam spot size. Therefore, the emittance growth is minimal when the beam is focused to a waist in the middle of the degrader. This is illustrated in Figure 3. In a variable thickness energy degrader the best results are obtained when the position of the degrader mid-point is constant and coincides with the beam focal point. Degraders with a flat front or exit face would require beam line retuning to move the beam focus according to the shift of the degrader center.

## 3. Energy degrader material

As we saw in the previous section, the main contribution to the beam emittance growth comes from the angular spread induced by the multiple scattering in the degrader material. The amount of angular scattering and, thus, the increase in the beam emittance depends on the degrader material. Here we will study how the multiple scattering angle varies for different materials when their thickness is set to provide the same amount of energy degradation. For proton beam energies below 250 MeV the Bethe-Bloch equation expressing material stopping power can be written ignoring the density correction term, which is important only above 1 GeV:

$$\frac{1}{\rho}\frac{dE}{dx} = K \frac{Z}{A} \left[ \frac{1}{\beta^2} \ln\left( \frac{2 m_e c^2 \beta^2 \gamma^2}{I(Z)} \right) - 1 \right] \qquad (6)$$

where $K = 0.307 \; MeV{\times}cm^2/g$. Expressed in the units of mass thickness, stopping power varies very little over a wide range of materials. The dependence on the material properties only comes in the atomic number to atomic weight ratio ($Z/A \approx 0.5$) and in a weak logarithmic dependence on the material's mean excitation energy $I(Z)$ which is tabulated for different materials in [7]. Thus, different energy degraders with the same mass thickness ($\rho dx$) will cause similar amount of energy degradation ($dE$).

Next, we estimate how the multiple scattering angle varies in materials of the same mass thickness. Equation (2) indicates that the multiple scattering angle is determined by the ratio of the material thickness to the material radiation length. The radiation length for different materials can be parameterized in the same units as mass thickness [5]:

$$\rho \cdot L0 = \frac{A}{Z} \cdot \frac{716.4 \ (g/cm^2)}{(Z+1) \cdot \ln(287/\sqrt{Z})} \tag{7}$$

Equation (7) clearly indicates that the low Z elements have larger radiation length and, therefore, smaller multiple scattering angle. In the following table we quote radiation length for different materials that could be used for degrading the beam energy. Note, that while lithium appears to have the largest radiation length, its usage is limited by very low melting temperature. Therefore, the best practical degrader material is beryllium.

**Table 1:** Radiation length (L0) and density for several possible degrader materials. For a complete table of materials see [5,6].

| Material | $\langle Z/A \rangle$ | $\rho$ (g/cm$^3$) | L0 (g/cm$^2$) |
|---|---|---|---|
| Li | 0.43221 | 0.534 | 82.76 |
| Be | 0.44384 | 1.848 | 65.19 |
| C | 0.49954 | 2.265 | 42.70 |
| Lexan | 0.52697 | 1.20 | 41.46 |
| Water | 0.55509 | 1.00 | 36.08 |
| Al | 0.48181 | 2.70 | 24.01 |

## 4. Energy degrader matching to the momentum selection line

Next, we discuss how the degrader geometry effects the beam emittance matching to the subsequent beam line. If orientation of the degraded beam emittance is tilted with respect to the beam line acceptance, then only a fraction of the beam will be transmitted through the beam line even if the areas are similar.

To investigate the problem we consider the beam emittance evolution through three different degrader configurations shown in Figure 1. The initial conditions are the same for all degrader configurations. We assume that the beam is focused to a waist at $z=0$. We then calculate the beam emittance as a function of the degrader thickness at $z=L$, which corresponds to the maximum degrader thickness. There are two effects that should be taken into account: 1) evolution of the beam emittance through drift space and 2) emittance growth in the degrader. Figure 4 illustrates evolution of the beam emittance through a drift space, while Figure 2 shows the effect of multiple scattering in a degrader to the beam emittance. The beam emittance starts with the upright orientation at the focal point ($z=0$). The drift space "tilts" the beam emittance in the phase space, while multiple scattering in the degrader "straitens" the beam emittance orientation. This qualitative analysis leads to a conclusion that drift space followed by energy degrader is preferable configuration to keep the beam emittance orientation constant at the entrance to the momentum selection beam line. This corresponds to the degrader geometry with a flat exit.

Next, we perform a more rigorous quantitative analysis. We will use the sigma transport matrix notation [8] to describe the beam emittance evolution.

$$\begin{aligned}
\sigma_{11} &= \varepsilon\beta = \langle x^2 \rangle; \\
\sigma_{12} &= -\varepsilon\alpha = -\langle x'x \rangle; \\
\sigma_{22} &= \varepsilon\gamma = \langle x'^2 \rangle
\end{aligned} \quad (8)$$

Then, the emittance evolution in a drift space of length *L* can be written as,

$$\begin{aligned}
\sigma_{11}(L) &= \sigma_{11}(0) + 2L \cdot \sigma_{12}(0) + L^2 \sigma_{22}(0); \\
\sigma_{12}(L) &= \sigma_{12}(0) + L \cdot \sigma_{22}(0); \\
\sigma_{22}(L) &= \sigma_{22}(0).
\end{aligned} \quad (9)$$

The multiple scattering contribution in the degrader can be written as:

$$\begin{aligned}
\sigma_{11} &= \sigma_{11} + \langle x_{MS}(L)^2 \rangle; \\
\sigma_{12} &= \sigma_{12} + \langle x_{MS}(L) \cdot \theta(L) \rangle; \\
\sigma_{22} &= \sigma_{22} + \langle \theta(L)^2 \rangle.
\end{aligned} \quad (10)$$

Combining the two effects together gives us complete description of the emittance evolution in different degrader geometries. The parameter of interest is the tilt angle of the emittance ellipsoid with respect to the coordinate system. For an ellipse parameterization shown in equation (4), the tilt angle can be written as:

$$\tan(2\phi) = \frac{2\alpha}{\gamma - \beta} = \frac{-2\sigma_{12}}{\sigma_{22} - \sigma_{11}} \tag{11}$$

One can easily check that a rotation of the phase space coordinate system (*x*, *x'*) by angle $\phi$ aligns the cardinal axes of the ellipse with the axes of the new reference frame. The ellipse parameterization in the new coordinates would simplify to

$$\varepsilon^2 = a^2 x_1^2 + b^2 x_1'^2 \tag{12}$$

We calculated the dependence of the emittance tilt angle on the energy degrader thickness for a beryllium energy degrader up to 15 cm thick. The results are plotted in Figure 5, for different degrader geometries. There are several conclusions to be made from the data. First, as expected from our qualitative analysis, the flat exit face geometry of the degrader provides the least amount of variation in the emittance orientation. This is due to the fact that the emittance evolution is dominated by the multiple angular scattering and emittance tilt in a drift space becomes significant only for small degrader thickness. Second, the emittance tilt angle is rather small (less than 150 mrad or 8.6 degree) in all degrader configurations provided that the degrader thickness exceeds 2 cm. For thinner degrader, the emittance is smaller than the acceptance of the beam line and the tilt in the emittance ellipse does not reduce the beam transmission efficiency. Taking into account additional beam scattering in the air and in the vacuum windows, we can conclude that geometry of a degrader does not make a significant impact on the beam line settings downstream of the degrader.

## 5. Conclusions

We can draw several conclusions from this study of performance optimization for a variable energy degrader. First, beryllium appears to be an optimal material for the energy degrader. For the same amount of energy degradation, beryllium causes the least amount of multiple scattering. Second, it appears to be very important to focus the incoming beam to a double waist in the middle of the degrader. Focusing the beam both vertically and horizontally minimizes the amount of emittance growth in any given energy degrader. Therefore, an optimal degrader performance would be achieved in a geometry with fixed mid-point (such as double wedge system). Finally, the emittance matching to the subsequent momentum selection beam line appears to be less critical, although a degrader with flat exit face offers somewhat better beam matching.

The results of this study have been integrated into the energy degrader design for the MPRI project [1]. As a part of the MPRI project, this study has been supported by a construction grant from the State of Indiana. The author is grateful to B. Gottschalk and N. Schreuder for raising questions that inspired this study.The results of this study have been integrated into the energy degrader design for the MPRI project [1]. As a part of the MPRI project, this study has been supported by a construction grant from the State of Indiana. The author is grateful to B. Gottschalk and N. Schreuder for raising questions that inspired this study.

## References

[1]  V.A. Anferov *et al*., "*The Indiana University Midwest Proton Radiation Institute*", in Proc. of 2001 Part. Accel. Conf. (**PAC-2001**), 645 (2001).

[2]  J.B. Flanz *et al*., "*Overview of the MGH-Northeast Proton Therapy Center plans and progress*", Nucl. Instrum. and Meth. B, **99**, 830-834 (1995).

[3]  W.T. Chu, B.A. Ludewigt, and T.R. Renner, "*Instrumentation for treatment of cancer using proton and light-ion beams*", Rev. Sci. Instrum. **64(8)**, 2055 (1993).

[4]  J.B. Flanz, F. Gerardi and E.L. Hubbard, "*Design considerations for a proton therapy beam line with an energy degrader*", Proc. of 14th Intl. Conf. on Appl. of Accel. in Research and Industry (**CAARI**), 1257 (1997).

[5]  D.E. Groom *et al*., "*Review of Particles Physics*", European Phys. Journal **C15**, 1 (2000).

[6]  Y.S. Tsai, Rev. Mod. Phys. **46**, 815 (1974).

[7]  "*Stopping Powers and Ranges for Protons and Alpha Particles*", ICRU Report #49, International Commission on Radiation Units and Measurements, (1993).

[8]  K. Brown *et al*., "*A First and Second Order Matrix Theory for the Design of Beam Transport Systems and Charged Particle Spectrometers*", SLAC-trans-075-rev4 (1982).

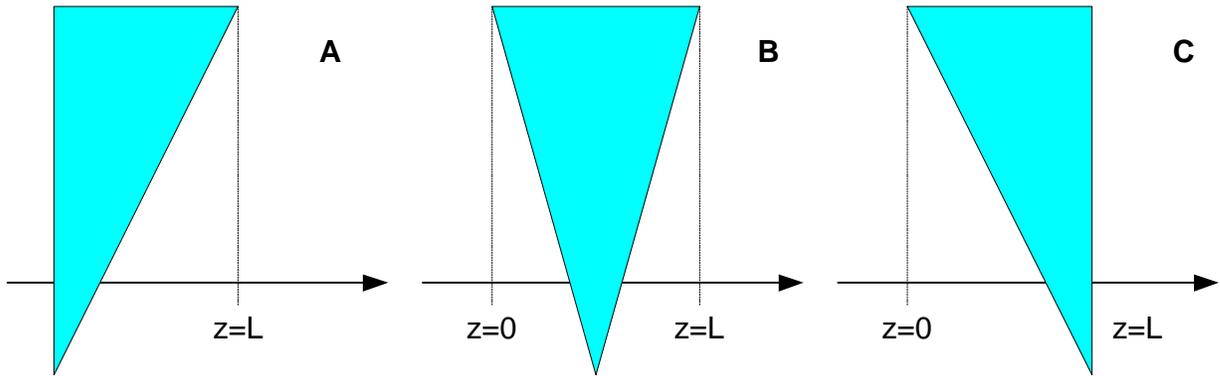

**Figure 1**: Different geometries of a variable energy degrader. (a) Flat front face, (b) fixed center of gravity and (c) flat exit face. Beam direction is from left to right.

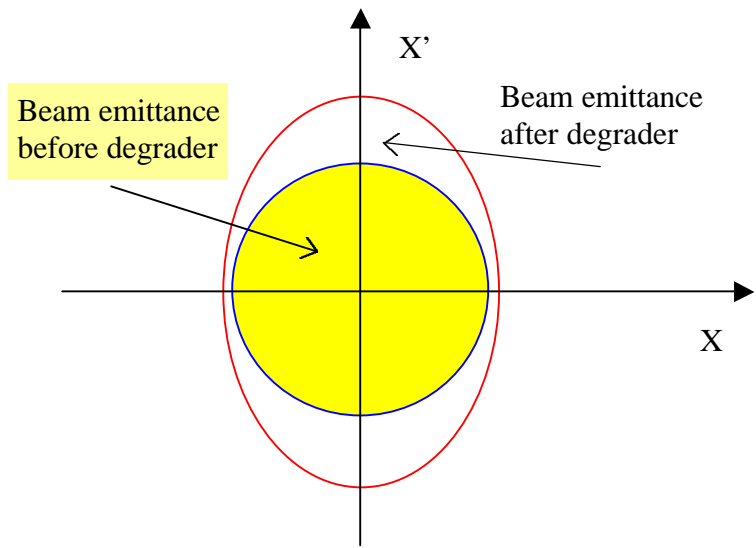

**Figure 2**: Effect of multiple scattering in a degrader on the beam emittance.

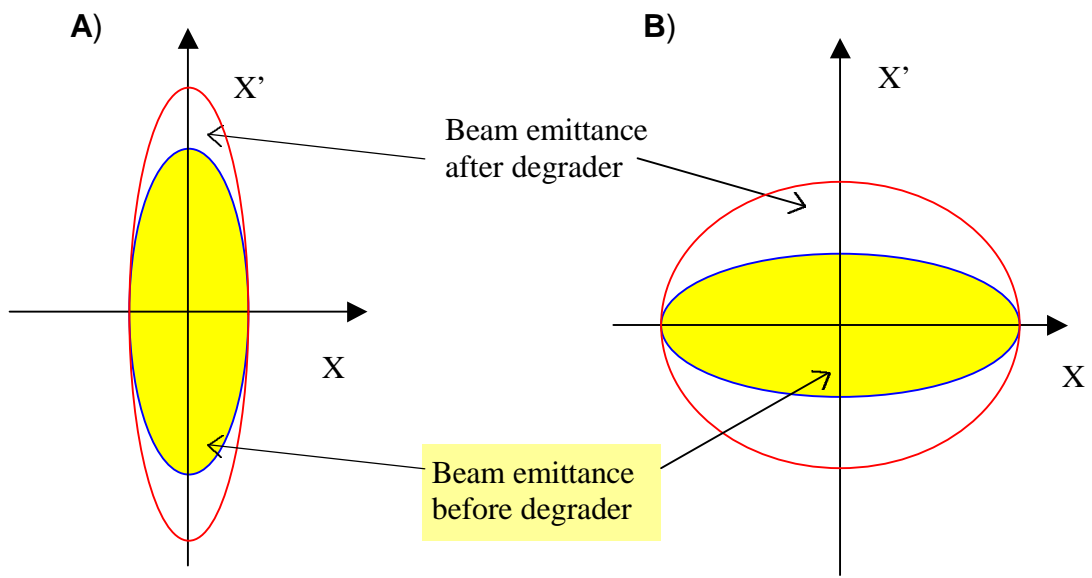

**Figure 3**: Beam emittance growth in a degrader under different focusing conditions.
(a) Beam goes through a waist at the degrader  (b) Beam is spread out at the degrader.

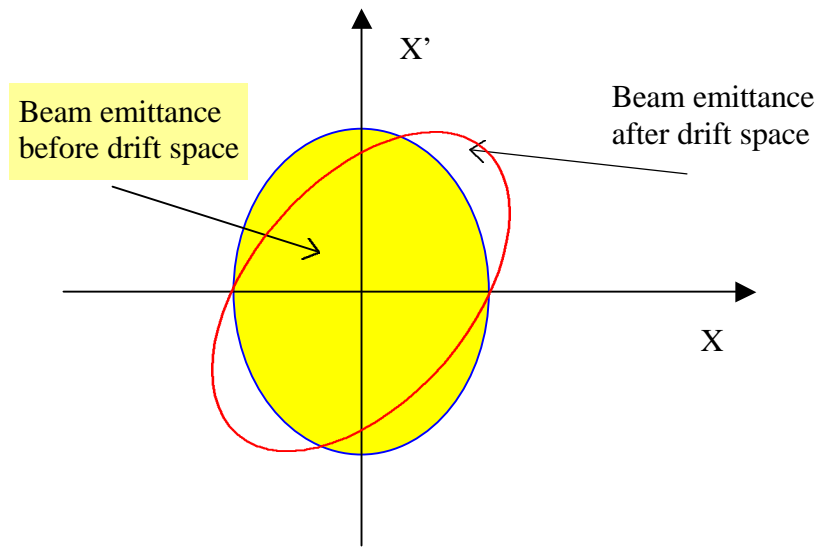

**Figure 4**: Evolution of the beam emittance through a drift space.

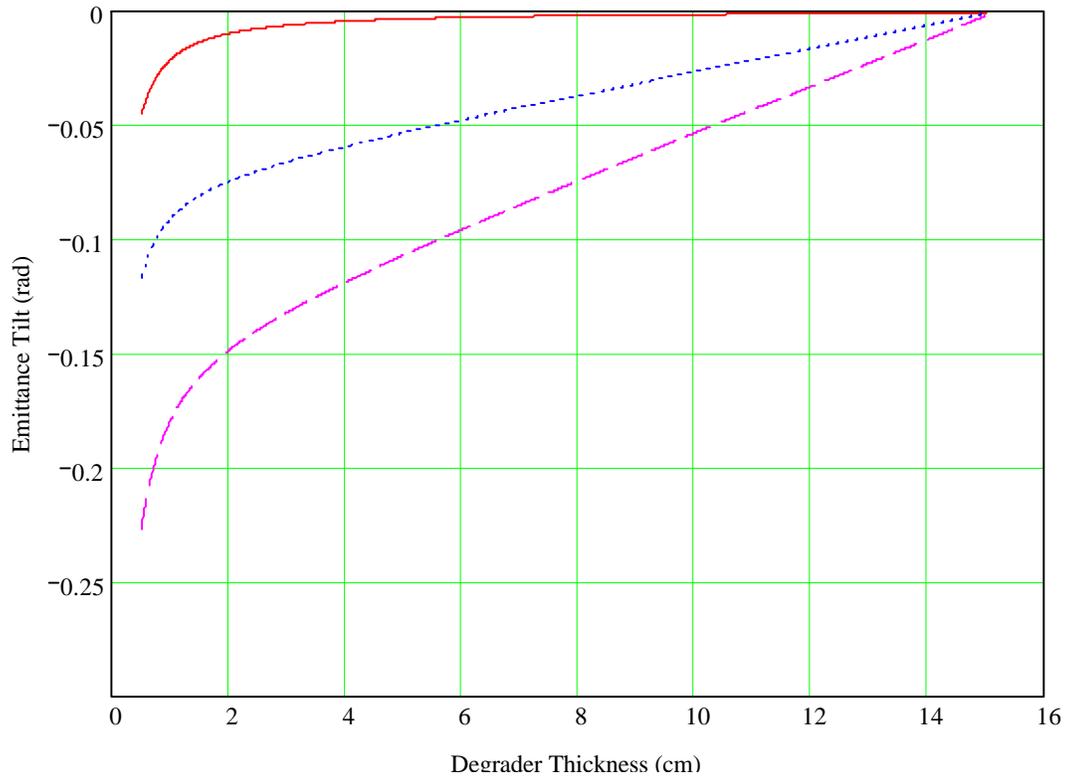

**Figure 5**: Beam emittance tilt is plotted versus the degrader thickness for different degrader geometries. Solid line corresponds to flat exit wedge, dotted line – to fixed center of gravity geometry, and dashed line – to flat front face.